\def\apj{{\em The Astrophysical Journal}}

\def\grl{{\em Geophys. Res. Lett.}}
\def\jgr{{\em J. Geophys. Res.}}

\def\nu{{\em Nuclear Fusion}}

\def\prl{{\em Phys. Rev. Lett.}}

\documentclass[%
 aip,groupedaddress,
 jmp,%
 amsmath,amssymb,
preprint,
floatfix,
]{revtex4-1}

\usepackage{graphicx}
\usepackage{float}
\usepackage{dcolumn}
\usepackage{bm}
\usepackage{amsmath,amsfonts,amsthm,amssymb}
\usepackage{xcolor}

\begin{document}
\preprint{APS/123-QED}
\title{ Electromagnetic Electron  Kelvin-Helmholtz Instability    }

\author{H. Che }
\email[Author to whom correspondence should be addressed: ]{hc0043@uah.edu}
\affiliation{Center for Space Plasma and Aeronomic Research (CSPAR), University of Alabama in Huntsville, Huntsville, AL 35805, USA}
\affiliation{Department of Space Science, University of Alabama in Huntsville, Huntsville, AL 35899, USA}
\author{G. P. Zank}
\affiliation{Center for Space Plasma and Aeronomic Research (CSPAR), University of Alabama in Huntsville, Huntsville, AL 35805, USA}
\affiliation{Department of Space Science, University of Alabama in Huntsville, Huntsville, AL 35899, USA}
\date{\today}

\begin{abstract}
On electron kinetic scales, ions and electrons decouple, and electron velocity shear on electron inertial length $\sim d_e$ can trigger electromagnetic (EM) electron Kelvin-Helmholtz instability (EKHI).  In this paper, we present an analytic study of EM  EKHI  in an inviscid collisionless plasma with a step-function electron shear flow.  We show that in incompressible collisionless plasma the  ideal  electron frozen-in condition $\mathbf{E} + \mathbf{v}_e \times \mathbf{B}/c = 0$ must be broken for the EM EKHI to occur. In a step-function electron shear flow, the  ideal   electron frozen-in condition is replaced by  magnetic flux conservation, i.e., $\nabla \times (\mathbf{E} + \mathbf{v}_e\times \mathbf{B}/c) = 0$,  resulting in a dispersion relation similar to that of the standard ideal and incompressible magnetohydrodynamics KHI. The magnetic field parallel to the electron streaming suppresses the EM EKHI due to magnetic tension. The threshold for the EM mode of the EKHI is $(\mathbf{k}\cdot\Delta\mathbf{U}_e)^2>\frac{n_{e1}+n_{e2}}{n_{e1} n_{e2}}[n_{e1}(\mathbf{v}_{Ae1}\cdot\mathbf{k})^2+n_{e2}(\mathbf{v}_{Ae2}\cdot\mathbf{k})^2]$, where $\mathbf{v}_{Ae} =\mathbf{B}/(4\pi m_e n_e)^{1/2}$, $\Delta\mathbf{U}_e$ and $n_e$ are the electron streaming velocity shear and densities, respectively. The growth rate of the EM mode is $\gamma_{em} \sim \Omega_{ce}$, the electron gyro-frequency. 
\end{abstract}

\maketitle
\section{Introduction}
The electromagnetic (EM)  Kelvin-Helmholtz instability (KHI) is one of the most common instabilities in nature. It is driven by  velocity shear in a single continuous fluid or a velocity difference across the interface between two fluids. Chandrasekhar's systematic studies\citep{chandra61book}  showed that the  KHI can also occur in incompressible magnetohydrodynamics, although for a magnetized plasma, magnetic tension parallel to the streaming can suppress the KHI. Subsequent investigations of KHI in  plasma were carried out in the compressible magnetohydrodynamics (MHD) framework \citep{dangelo65pof,perkins71pof,lau80pof,miura82prl,wang92pof}. Typically, KHI has been considered as a large-scale fluid instability and its importance on kinetic scales has not been appreciated until recently.

Recent observations and kinetic simulations have found that KHI on kinetic scales plays an important role in electron acceleration in explosive events, such as planetary magnetospheric substorms and solar flares \citep{henri13pop,lilje14jgr,huangC17apj,zhong18prl,hwang19grl,che21apjb,che21apja, dela21jgr,che20apj,li22natcommu}. The instability is driven largely by the shear in electron streams and hence is called an electron Kelvin-Helmholtz instability (EKHI). In magnetic reconnection in collisionless plasma, the current sheet shrinks to a width close to the electron inertial length $d_e$ before triggering explosive reconnection events. The EKHI is common in magnetic reconnections that have  velocity shear due to the anti-parallel electron streaming along the magnetic field lines in a manner similar to the MHD  KHI \citep{che20apj}, but the growth rate is much higher than that of MHD KHI and the wavelength is much shorter.  Interestingly, the \textit{Magnetospheric Multiscale} (MMS) observations appears to have discovered  EKHI and the corresponding vortices \citep{wilder16grl,zhong18prl,hwang19grl,wilder20grl,li22natcommu}.

 The EKHI has not been studied analytically as a distinctively different instability from the MHD KHI, even for the simplest case.  The simplest and most commonly cited case is the EM KHI in an inviscid and incompressible fluid for a step function velocity shear flow\citep{chandra61book}.  If  the velocity $\mathbf{U}_1$ parallel to the magnetic field $\mathbf{B}_1$ and $\mathbf{U}_2$ parallel to the magnetic field $\mathbf{B}_2$ are separated by an interface $z_s$ (Fig.\ref{shear}), the dispersion relation is
\begin{gather}
 \omega = \frac{\rho_1(\mathbf{k}\cdot\mathbf{U}_1)+\rho_2(\mathbf{k}\cdot\mathbf{U}_2) }{\rho_1+\rho_2} 
\pm \frac{i}{\rho_1+\rho_2} \Xi^{1/2} \label{chandra}; \\
\Xi = \rho_1\rho_2(\Delta\mathbf{U}\cdot\mathbf{k})^2-(\rho_1+\rho_2)(n_1(\mathbf{v}_{A1}\cdot\mathbf{k})^2+n_2(\mathbf{v}_{A2}\cdot\mathbf{k})^2) , \nonumber
\end{gather}
where $\Delta\mathbf{U} \equiv \mathbf{U}_1 - \mathbf{U}_2$, $v_{A} \equiv B/(4\pi \rho)^{1/2}$ is the  Alfv\'en velocity and $\rho= m_i n_i + m_e n_e$. We can see that a magnetic field parallel to the flow direction suppresses KHI. 


 On electron dynamic scales $\sim d_e$, ions and electrons decouple. Ions are demagnetized and can be treated as a background, and electron dynamics dominates. At MHD scales the Ohm's law and the momentum equation are two independent equations, but on electron scales the electron momentum equation is also the Ohm's law \citep{birn01jgrb},  illustrating the key difference between the MHD scale and electron scale KHI: the equations  that govern the fluid dynamics on different scales do not have an one-to-one correspondence. On MHD scales, the ideal Ohm's law/frozen-in condition and is typically assumed to derive the standard KHI dispersion relations \citep{chandra61book}, including the one shown in Eq.~(\ref{chandra}). On electron scales, the overlap of the momentum equation and Ohm's law makes it unclear whether we can simply extend the dispersion relation for ideal MHD EHI to EKHI, in particular, to Eq.~(\ref{chandra}).

 In the paper we derive the dispersion relations for  the EM mode of the EKHI in inviscid and incompressible collisionless plasma with a step function velocity shear flow. 
We show in incompressible and inviscid collsionless plasma,
the  ideal  frozen-in condition $\mathbf{E} + \mathbf{v}_e\times \mathbf{B}/c = 0$ must be broken for EM EKHI to occur regardless of the functional form of the velocity shear.The reason is that the frozen-in condition decouples the electron dynamics from the magnetic and electric fields.  In an incompressible step function  electron velocity shear flow, the frozen-in condition is replaced by magnetic flux conservation  $\nabla \times (\mathbf{E} + \mathbf{v}_e\times \mathbf{B}/c) = 0$. In this case, the magnetic field plays a similar role to that in Eq.~(\ref{chandra}) and the electron  Alfv\'en velocity $\mathbf{v}_{Ae} =\mathbf{B}/(4\pi m_e n_e)^{1/2}$ replaces the role of the MHD Alfv\'en velocity $v_A$. The threshold for the EM EKHI to occur is $(\mathbf{k}\cdot\Delta\mathbf{U}_e)^2>\frac{n_{e1}+n_{e2}}{n_{e1} n_{e2}}[n_{e1}(\mathbf{v}_{Ae1}\cdot\mathbf{k})^2+n_2(\mathbf{v}_{Ae2}\cdot\mathbf{k})^2]$, where $\Delta\mathbf{U}_e$ and $n_e$ are the electron streaming velocity shear and densities, respectively. The growth rate is $\sim \Omega_{ce}$, the electron gyro-frequency.

\section{ Electron Dynamic Equations and The Step Function Electron  Shear Flow }
\label{sec2}
On electron dynamic scales ranging from the electron Debye length $\lambda_{De} \equiv v_{te}/\omega_{pe}$ to the electron inertial length $d_e \equiv c/\omega_{pe}$, where  $v_{te}$ is the electron thermal speed and $\omega_{pe}$ is the electron plasma frequency, electrons and ions are no longer strongly coupled. Electrons dominate the high-frequency dynamics and ions behave like  a stationary background  due to significantly larger mass. As a result, electrons carry most of the current and are responsible for charge separation. Thus on electron dynamical scales we neglect the ion dynamics. We assume that the plasma is inviscid  and incompressible collisionless and the electron fluid equations are 
\begin{gather}
  \partial_t n_e + \nabla\cdot(n_e\mathbf{v}_e) =  0;\label{con}\\
  m_e n_e (\partial_t + \mathbf{v}_e\cdot\nabla) \mathbf{v}_e + e n_e (\mathbf{E} +\frac{ \mathbf{v}_e}{c} \times \mathbf{B}) + \nabla P_e  = 0; \label{mom}\\
  \nabla \cdot \mathbf{v}_e = 0, \label{divv}
\end{gather}
and we assume  that the electron pressure is a scalar $P_e$, the system is coupled  with Maxwell equations,
\begin{eqnarray}
\nabla\cdot \mathbf{E}&=& 4\pi e (n_i -n_e); \label{possion}\\
\nabla\cdot \mathbf{B} &=& 0;\label{divb}\\
\nabla \times  \mathbf{E }&=&-\frac{1}{c}\partial_t \mathbf{B};\label{far}\\
\nabla \times \mathbf{B} &=&  -\frac{4\pi e n_e}{c} \mathbf{v}_e,\label{am}
\end{eqnarray}
where we neglect the temporal variation of the electric field in Ampere's law. The spatial scale of the inductive electric field is $\sim d_e$ the electron inertial length, and the time variation of the magnetic field is $\sim \Omega_{ce}\equiv eB/m_ec$ the electron gyro-frequency, thus in the Faraday's law $(\Delta E/\Delta t )/ (\Delta B/\Delta t) \sim v_{Ae}/c   \ll 1$, where $\mathbf{v}_{Ae}\equiv \mathbf{B} /(4\pi m_e n_e)^{1/2}$ is the electron Alfv\'en wave speed. Consequently the electric displacement $\Delta E/\Delta t$ is negligible compared to the electron current density in Ampere's law.  The current density on electron kinetic scales can be approximated as $\mathbf{j} \approx \mathbf{j}_e = -en_e\mathbf{v}_e$ and the ions' contribution in Ampere's equation (\ref{am}) is neglected.  

\begin{figure}[t]
\includegraphics[scale=2,trim=0 0 0 0,clip]{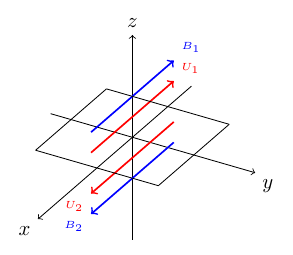} 
\caption {An illustration of the coordinate system. The two uniform magnetic fields  $\mathbf{B}_1$ and $\mathbf{B}_2$ are aligned along the stream velocity, and their directions can be ether parallel or anti-parallel. We will show that the EKHI depends only on the square of the magnetic field and is independent of their directions.}
\label{shear}
\end{figure}

Following Chandrasekhar\citep{chandra61book},   we explore both the EM mode EKHI for the simplest case as shown in Fig.~\ref{shear}: two uniform  electron fluids in relative horizontal motion along $x$ separated by a horizontal boundary at $z=0$  where the electron velocities $\mathbf{U}_1$ and $\mathbf{U}_2$ are discontinuous. 
\begin{equation}
\mathbf{U}_0 = 
\begin{cases}
U_1 \hat{x}, \quad z > 0\\
U_2 \hat{x}, \quad z < 0
\end{cases}
\end{equation}

On both sides of the boundary we assume the plasma is neutral but with different plasma densities. We assume the initial electron density is $n_1$ and $n_2$, 
\begin{equation}
n_{e0} = 
\begin{cases}
n_1, \quad z > 0\\
n_2, \quad z < 0
\end{cases}
\end{equation}
The initial electric field $\mathbf{E}_0 =0 $, since the initial densities of electrons and ions are equal, i.e. $n_{e0}=n_{i0}$. 

 The magnetic field $\mathbf{B}_1$ and $\mathbf{B}_2$ are uniform. In particular,  we will show that both the dispersion relations of EM in such a velocity shear configuration are independent of whether $\mathbf{B}_1$ and $\mathbf{B}_2$ are parallel or anti-parallel. 
\begin{equation}
\mathbf{B}_0 = 
\begin{cases}
B_1 \hat{x}, \quad z > 0\\
B_2 \hat{x}, \quad z < 0
\end{cases}
\end{equation}

With this configuration, an out-of-plane component of the magnetic field $B_y$ is also present, which is proportional to $z$, i.e.  $B_y \varpropto z$. When approaching the boundary interface $z=0$, $B_y$ approaches zero. We neglect the $B_y$ component in this study since the EKHI occurs in the neighborhood of  $z=0$.  The initial pressure $P_0$ is determined by the initial equilibrium   $P_0 + B_0^2/8\pi = constant$  on both sides of $z=0$. 


\section{ Electromagnetic Electron Kelvin-Helmholtz Instability}

\subsection{The Breaking of the  ideal   Frozen-in Condition  in EM EKHI in  Incompressible Plasma }
In the following we show that in the incompressible plasma, the  ideal   electron frozen-in condition $\mathbf{E} + \mathbf{v}_e  \times \mathbf{B}/c = 0 $ must be broken for EM EKHI to occur.  

 When the electrons are frozen-in with the magnetic field,  the electron momentum equation (\ref{mom}) reduces to:  
\begin{equation}
 \mathbf{E} + \mathbf{v}_e  \times \mathbf{B}/c = 0;
 \label{frozen-con}
\end{equation}
%
The linearization of  Eq.~(\ref{possion}) - ~(\ref{am}) gives 
\begin{equation}
\nabla^2 \delta \mathbf{E} + \nabla \nabla \cdot \mathbf{\delta E}=-\frac{4\pi }{c^2} \partial_t \delta \mathbf{j}_e,
\label{ddE}
\end{equation}
The EM EKHI wavelength is $\sim d_e$, thus Eq.~(\ref{ddE}) gives 
\begin{equation}
\omega_{pe}^2 \delta E \sim 2\pi\partial_t \delta j_e.
\label{ddEappro}
\end{equation} 

From the electron momentum equation (\ref{mom}), we can see that one of the requirements for the  ideal  electron frozen-in condition to be satisfied is
\begin{equation}
 m_e n_{e0} (\partial_t \delta v_e + \mathbf{v}_e\cdot \nabla \delta v_e) \ll n_{e0} e \delta E,
 \end{equation}
 This leads to
 \begin{equation}
  2\pi \partial_t \delta j_e \ll \omega_{pe}^2 \delta E.
  \end{equation}
This result contradicts equation (\ref{ddEappro}).  Therefore, in incompressible plasma, the  ideal   electron frozen-in condition prevents the EM EKHI from occurring. This conclusion is consistent with the fact that the frozen-in condition decouples the magnetic field from the electron fluid dynamics. This decoupling occurs because the frozen-in condition separates the electron momentum equation from the magnetic and electric fields. As a result, the growth rate of the instability is only determined by the electron velocity shear, similar to the fluid Kelvin-Helmholtz Instability. This is due to the fact that the electron Ohm's law is equivalent to the momentum equation.

\subsection{The Threshold and Growth Rate of EM EKHI in a Step Function Incompressible Electron Shear Flow}

In the following, we neglect the subscript $e$ for electrons. 
Here we derive the dispersion relation for EM EKHI in step function incompressible electron shear flows as shown in the Fig.~(\ref{shear}). 


 In Fig.~(\ref{shear}), the initial velocity shear is uniform at the both sides of the interface $z = 0$ and is symmetric to the out-of-plane $y$-direction, thus this is a two-dimensional flow problem in inviscid plasma, the vorticity is  conserved.  The initial vorticity is zero and is conserved at $z\neq 0$, i.e.
\begin{equation}
\nabla \times  \mathbf{v} =0 .
\label{inv}
\end{equation}
  Incompressibility  $\nabla\cdot \mathbf{v}=0$ and Eq.(\ref{inv})  at $z \neq 0$ give 
\begin{equation}
\nabla^2  \mathbf{v} =0.
\label{wavefun}
\end{equation}
From the above analysis, we can see that  for the EM mode, the continuity equation (\ref{con}) and Poisson's equation (\ref{possion}) are replaced by the Laplace equation Eq.~(\ref{wavefun}).

The magnetic field-induced electric field $\mathbf{E}$  impacts the electron momentum equation (\ref{mom}) through Faraday's law, i.e., Eq.~(\ref{far}). Taking the curl of Eq.~(\ref{mom}),  at $z\neq 0$ we obtain 
\begin{equation}
\nabla \times (\mathbf{E}+ \frac{\mathbf{v}}{c}\times \mathbf{B}) = 0.
\label{openfro}
\end{equation}
However, we can check that Eq.(\ref{openfro}) is also satisfied at $z = 0$.

Equation (\ref{openfro}) can be considered as an extension of the  ideal  electron frozen-in condition in an inviscid and incompressible electron fluid. The  ideal  electron frozen-in condition $\mathbf{E}+ \frac{\mathbf{v}}{c}\times \mathbf{B} = 0$ is the simplest (trivial) case of Eq.~(\ref{openfro}). In incompressible plasma, the EM EKHI requires the breaking of the  ideal   electron frozen-in condition  and  magnetic flux conservation is the simplest replacement. 

Using Faraday's law, $\nabla\cdot \mathbf{B}=0$ and  $\nabla\cdot \mathbf{v}=0$, we rewrite Eq.~(\ref{openfro}) as
\begin{equation}
\partial_t \mathbf{B} = \mathbf{B}\cdot\nabla\mathbf{v} - \mathbf{v}\cdot\nabla\mathbf{B}.
\label{bv}
\end{equation}
Eq.~(\ref{bv}) connects the velocity and magnetic field. We can rewrite the electron momentum equation (\ref{mom}) using Ampere's Law, i.e., Eq.~(\ref{am}) to obtain
\begin{equation}
mn(\partial_t +\mathbf{v}\cdot\nabla)\mathbf{v} +en\mathbf{E} -\frac{1}{4\pi} \mathbf{B}\cdot\nabla\mathbf{B} + \nabla P^{\ast} = 0,
\label{newmom}
\end{equation}
where 
\begin{equation}
P^{\ast} = P + \frac{B^2}{8\pi}
\end{equation}
 is the total pressure including magnetic pressure. 
 
 Eqs.~(\ref{wavefun}), (\ref{bv}), and (\ref{newmom}) form a complete set of equations that describe the EM mode of EKHI in incompressible and inviscid plasma for the configuration in Fig.~\ref{shear}. We linearize these equations to obtain
 \begin{gather}
\nabla^2 \delta \mathbf{v} = 0;\label{d2}\\
\partial_t \delta \mathbf{B} - \mathbf{B}_0\cdot \nabla \delta \mathbf{v} + \mathbf{U}_0 \cdot \nabla \delta \mathbf{B} + \delta \mathbf{ v}\cdot \nabla \mathbf{B}_0 -\delta \mathbf{B}\cdot \mathbf{U}_0=  0; \label{dbv}\\
mn_0 (\partial_t + \mathbf{U}_0\cdot \nabla)\delta \mathbf{v} + mn_0\delta\mathbf{v} \cdot \nabla \mathbf{U}_0 + en_0 \delta \mathbf{E}  -\frac{1}{4\pi}  \mathbf{B}_0\cdot \nabla\delta\mathbf{B} - \frac{1}{4\pi} \delta \mathbf{B}\cdot \nabla \mathbf{B}_0 + \nabla \delta P^{\ast} = 0 \label{dmom},
\end{gather}
where we have used $\mathbf{v} = \mathbf{U}_0 + \delta \mathbf{v}$ and $n= n_0 + \delta n$; $\mathbf{U}_0$ and $n_0$ represent the initial velocity $\mathbf{U}_1$ or $\mathbf{U}_2$ and density $n_1$ or $n_2$ respectively.  We also used $\mathbf{E} = \delta \mathbf{E}$ due to $\mathbf{E}_0 = 0$, and $\mathbf{B} = \mathbf{B}_0 + \delta \mathbf{B}$  and $P^{\ast} = P^{\ast}_0 + \delta P^{\ast}$, and 
\begin{gather}
P_0^{\ast}  = P_0 + \frac{B_0^2}{8\pi},\\
\delta P^{\ast} = \delta P + \frac{\mathbf{B}_0\cdot \delta \mathbf{B}}{4\pi}.
\end{gather}
$\mathbf{B}_0$ and $P_0$ represent the initial magnetic field and pressure, respectively. $\mathbf{E}_0 = 0$ leads to the term $\mathbf{E}_0 \delta n $ vanishes and as a result, the density discontinuity at $z=0$ does not affect the linearization. 

Perturbations are continuous in the $xy$ plane and discontinuous at $z = 0$ in the $z$ direction. We consider perturbations of the form $\delta f(z) e^{i(k_x  x +k_y y -\omega t)}$. As we demonstrated in Section II, the ratio $(\Delta E/\Delta t )/ (\Delta B/\Delta t) \sim v_{Ae}/c \ll 1$, indicating that the influence of the induced electric field is insignificant.  As a result, the induced electric field is not a determining factor in the growth rate of the EKHI. Since we neglected $\partial_t\mathbf{E}$ in Ampere's law, the induced electric field  is also neglected when we take the time derivative of the linearized equations (\ref{dmom}). Then the equations (\ref{d2}) and (\ref{dbv}), and the time derivation of equation (\ref{dmom}) can be approximated as follows:

\noindent The velocity Laplace equation is valid at $z \neq 0$,
 
\begin{equation}
(\partial_z^2 - k^2) \delta \mathbf{v} = 0, 
\label{d2f}
\end{equation}

\noindent and the magnetic flux conservation and the momentum equation are valid in the whole space

\begin{gather}
-i\Omega \delta \mathbf{B} - i\mathbf{B}_0\cdot \mathbf{k} \delta \mathbf{v} + \delta \mathbf{v}\cdot \nabla \mathbf{B}_0 - \delta \mathbf{B} \cdot \nabla \mathbf{U}_0 =  0; \label{dbvf}\\
-i m n_0 \Omega \delta \mathbf{v} + mn_0 \delta \mathbf{v} \cdot \nabla \mathbf{U}_0
- \frac{i}{4\pi}\mathbf{B}_0\cdot \mathbf{k}  \delta \mathbf{B} -\frac{1}{4\pi}\delta \mathbf{B}\cdot \nabla \mathbf{B}_0+\nabla  \delta P^{\ast}  = 0,\label{dmomf}
\end{gather}
where $\Omega = \omega - \mathbf{k}\cdot \mathbf{U}_0$ and $\mathbf{k} = k_x \hat{x} + k_y \hat{y}$. 

The component equations of Eq.(\ref{dmomf}) are 
\begin{gather}
mn_0 \partial_t \delta v_x + mn_0 U_0\partial_x \delta v_x +mn_0\delta v_z\partial_z U_0 -\frac{1}{4\pi}B_0\partial_x\delta B_x -\frac{1}{4\pi}\delta B_z \partial_z B_0 + \partial_x P^{\ast} = 0; \label{xcom}\\
mn_0 \partial_t \delta v_y + mn_0 U_0 \partial_x \delta v_y -\frac{1}{4\pi} B_0\partial B_y + \partial_y P^{\ast} =0; \label{ycom}\\
mn_0 \partial_t \delta v_z + mn_0 U_0 \partial_x \delta v_z -\frac{1}{4\pi} B_0\partial B_z + \partial_z P^{\ast} =0. \label{zcom}
\end{gather} 

We multiply both sides of Eq. (\ref{xcom}) and  Eq. (\ref{ycom}) by $i k_x$  and $i k_y$ respectively, then sum, and  make use of $\nabla\cdot\delta \mathbf{v} =0$ and  $\nabla\cdot\delta \mathbf{B} =0$ to obtain

\begin{equation}
k^2 P^{\ast} = imn_0 \Omega \partial_z \delta v_z + i\frac{\mathbf{k}\cdot \mathbf{B}_0}{4\pi} \partial_z \delta B_z+ ik_x m n_0 \partial_z U_0 \delta v_z-i \frac{k_x}{4\pi}\partial_z B_0 \delta B_z.
\label{pstar}
\end{equation} 

The $z$ component of Eq.(\ref{dbvf}) gives
\begin{equation}
\delta B_z = - \frac{k_x B_0}{\Omega}\delta v_z.
\label{bzvz}
\end{equation} 

On inserting Eq.(\ref{pstar}) into Eq.(\ref{zcom}) and using Eq.(\ref{bzvz}), we obtain
\begin{multline}
- m n_0 k^2 \Omega \delta v_z + \frac{k^2 k_x^2 B_0^2}{4\pi \Omega}\delta v_z 
+ \partial_z [ mn_0 \Omega \partial_z \delta v_z
 - \frac{\mathbf{k}\cdot{\mathbf{B}}_0}{4\pi}\partial_z (\frac{\mathbf{k}\cdot \mathbf{B}_0}{\Omega} \delta v_z)\\+k_x mn_0 \partial_z U_0 \delta v_z 
 +\frac{\mathbf{k}\cdot{\mathbf{B}}_0}{4\pi \Omega}\partial_z (\mathbf{k}\cdot \mathbf{B}_0) \delta v_z]=0.
 \label{vzeq}
 \end{multline}

Eq.~(\ref{d2f}) together with the boundary conditions $\delta v_z=0$ at $z=\infty$  gives the general solution 
\begin{eqnarray}
\delta v_z \varpropto e^{i(k_x  x +k_y y -\omega t)}e^{-kz},  \qquad z>0,\\
\delta v_z \varpropto e^{i(k_x  x +k_y y -\omega t)}e^{kz},  \qquad z<0.
\end{eqnarray}
We can then write $\delta f(z)$ as a function proportional to $e^{-k\vert z\vert}$.  $\delta \mathbf{v}$ is not continuous across $z=0$. But $\delta \mathbf{v} =d\delta \mathbf{l}/dt = -i \Omega \delta \mathbf{l}$,  where $\delta \mathbf{l}$ is the displacement of the any fluid element on the interface which  is continuous at the $z=0$ plane, thus we can rewrite $\delta \mathbf{v}$ as
\begin{equation}
\delta \mathbf{v} = \delta \mathbf{v}_0 \Omega e^{-k\vert z\vert},
\end{equation}
where $\delta \mathbf{v}_0$ is the velocity perturbation at $z=0$.

On Integrating Eq. (\ref{vzeq}) over the interface $z=0$ from  $0-\epsilon $ to $0+\epsilon$, where $\epsilon\rightarrow 0 $ and  applying the solution $\delta v_z$ for the two regions, we obtain 
\begin{equation}
n_1 (\Omega_1^2 - (\mathbf{v}_{Ae1}\cdot\mathbf{k})^2) + n_2 (\Omega_2^2 - (\mathbf{v}_{Ae2}\cdot\mathbf{k})^2) = 0,
\label{emdisp1}
\end{equation}
where $\mathbf{v}_{Ae1} = (B_1^2/4\pi m n_1)^{1/2}\mathbf{B}_1/B_1$, $\mathbf{v}_{Ae1} = (B_2^2/4\pi m n_2)^{1/2} \mathbf{B}_2/B_2$, 
$\Omega_1 = \omega - \mathbf{k}\cdot \mathbf{U}_1$ and $\Omega_2= \omega - \mathbf{k}\cdot \mathbf{U}_2$.
On rearranging Eq.(\ref{emdisp1}), we obtain the dispersion relation for  the  EM mode of  EKHI as
\begin{equation}
 \omega = \frac{n_1(\mathbf{k}\cdot\mathbf{U}_1)+n_2(\mathbf{k}\cdot\mathbf{U}_2) }{n_1+n_2} 
\pm \frac{i}{n_1+n_2}[n_1n_2(\Delta\mathbf{U}\cdot\mathbf{k})^2-(n_1+n_2)(n_1(\mathbf{v}_{Ae1}\cdot\mathbf{k})^2+n_2(\mathbf{v}_{Ae2}\cdot\mathbf{k})^2)]^{1/2},
\label{emdisp}
\end{equation}
where $\Delta\mathbf{U}=  \mathbf{U}_1 - \mathbf{U}_2$.
We can see that the threshold for the occurrence of EM mode in the EKHI is 
\begin{equation}
(\mathbf{k}\cdot\Delta\mathbf{U})^2>\frac{n_1+n_2}{n_1 n_2}[n_1(\mathbf{v}_{Ae1}\cdot\mathbf{k})^2+n_2(\mathbf{v}_{Ae2}\cdot\mathbf{k})^2].
\label{thresh1}
\end{equation}

 For $U \gg v_{Ae}$,  we can neglect the magnetic field, then for $\Delta U \sim v_{Ae}$, and $1/k \sim d_e$, the growth rate $\gamma_{em}$ is about $\gamma_{em}\sim \Omega_{ce}$,  is the electron gyro-frequency.  For a special but common case where $n_1=n_2$ and $B_1=B_2$,  the threshold is 
 \begin{equation}
 \Delta U> 2v_{Ae} .
 \label{thresh2}
 \end{equation}
 
 If the electron velocities on the two sides are antiparallel $\mathbf{U}_1=-\mathbf{U}_2$, the real frequency of the EKHI wave $\omega_r$ is zero, i.e. $\omega_r =0$ and $U>v_{Ae}$.

Comparing the dispersion relations of the EM mode of the EKHI in Eq.~(\ref{emdisp}) and the dispersion relation of the ideal incompressible MHD KHI in Eq.~(\ref{chandra}),  we find that Eq.~(\ref{emdisp}) is the same as Eq.~(\ref{chandra}) if we replace the electron Alfve\'n wave speed $v_{Ae}$ with the Alfve\'n wave speed $v_A$. In both cases, magnetic tension resists the development of the instability and the pressure and magnetic pressure supports the formation of vortices.  However, the EM mode of the EKHI is not a direct extension of ideal incompressible MHD KHI, in that the dynamics is completely different. On the incompressible MHD scale, the plasma is frozen-in with the magnetic field $\mathbf{E} + \mathbf{v}\times \mathbf{B}/c=0$ where $\mathbf{v}$ is the MHD velocity. However, on the electron dynamic scale, the frozen-in condition must be broken for the EM EKHI to occur in incompressible plasma. In the simplest case magnetic flux conservation $\nabla \times(\mathbf{E} + \mathbf{v}_e\times \mathbf{B}/c)=0$ replaces $\mathbf{E} + \mathbf{v}_e\times \mathbf{B}/c=0$. This condition allows both the occurrence of electron heating and electron motions that are not completely decoupled from the magnetic field, although constrained, and explains why the magnetic field increases the threshold of the EM mode of the EKHI, in particular a uniform magnetic field parallel to the direction of electron streaming suppresses the EM mode of the EKHI thanks to the magnetic tension. 
  
\section{Discussion and Conclusions}
\label{disc}
In this paper, we have investigated the threshold criteria and growth rates of the electromagnetic (EM)  electron Kelvin-Helmholtz instabilities for step function velocity shear flows using an electron fluid model coupled with Maxwell's equations in an inviscid and collisionless plasma. Unlike the KHI in ideal incompressible magnetohydrodynamics (MHD), we show that the  ideal  electron frozen-in condition must be broken for the EM EKHI to occur in incompressible plasma. Similar to the EM KHI in incompressible step function velocity shear flows,  the magnetic tension parallel to the velocity shear inhibits  the development of the EM EKHI and thus the electron fluid velocity shear must be larger than the electron Alfv\'en speed, i.e., $\Delta U> v_{Ae}$ to trigger the instability. The wavelength of the EM mode of the EKHI is of the order of the electron inertial length $d_e$ and the growth rate is of the order of the electron gyro-frequency $\gamma_{em} \sim \Omega_{ce}$. 

 The dispersion relations for the EM  mode and the relevant thresholds criteria and growth rates can be summarized as  follows.

\noindent \textit{General Dispersion Relation}:
  \begin{gather}
 \omega  = \omega_r \pm i \gamma_{em} , \nonumber\\
 \omega_r =  \frac{n_1(\mathbf{k}\cdot\mathbf{U}_1)+n_2(\mathbf{k}\cdot\mathbf{U}_2) }{n_1+n_2}, \nonumber\\
\gamma_{em} = \frac{1}{n_1+n_2}[n_1n_2(\Delta\mathbf{U}\cdot\mathbf{k})^2-(n_1+n_2)(n_1(\mathbf{v}_{Ae1}\cdot\mathbf{k})^2+n_2(\mathbf{v}_{Ae2}\cdot\mathbf{k})^2)]^{1/2}. \nonumber
  \end{gather}
\noindent \textit{Threshold}:
\begin{equation}
(\mathbf{k}\cdot\Delta\mathbf{U})^2>\frac{n_1+n_2}{n_1 n_2}[n_1(\mathbf{v}_{Ae1}\cdot\mathbf{k})^2+n_2(\mathbf{v}_{Ae2}\cdot\mathbf{k})^2]. \nonumber
\end{equation}

  For a simple but common case where $n_1=n_2$ and $B_1=B_2$,  the threshold becomes
 \begin{equation}
 \Delta U> 2v_{Ae} . \nonumber
 \label{uae}
 \end{equation}
 
  If the velocity shear is antiparallel, i.e., 
  $
  \mathbf{U}_1=-\mathbf{U}_2, \nonumber
  $
   the real frequency of the EKHI wave is 
   \begin{equation}
   \omega_r =0. \nonumber
    \end{equation}
    
 We have presented an electron fluid analytic solution for EKHI, which is more general than the qualitative result presented by \citet{Fermo12prl}. In their study, Fermo et al. estimated the threshold for EKHI by assuming that the growth rate of EKHI for a wavenumber of $k\sim1/d_e$ is approximately $\gamma_{em}\sim\Delta U/d_e$. This estimate is an extension from the growth rate of KHI for weak magnetic fields in uniform plasma. Additionally, they assumed that the growth rate of EKHI should exceed the whistler frequency. With these assumptions, they obtained the threshold for EKHI as $\Delta U>v_{Ae}/2$, with the growth rate $\gamma_{em}=\Omega_{ce}/2$ - which is an estimate for the special case of $\gamma_{em}$ that we presented above.
  
For a weak magnetic field and a uniform plasma density, our result allows us to approximate $\gamma_{em}=\Delta U/(2k)$. In this case, the threshold for EKHI to occur is theoretically $\Delta U>0$. However, if we consider the growth rate to be $\gamma_{em}\sim\Omega_{ce}/2$ and $1/k\sim d_e$, we obtain the threshold $\Delta U=v_{Ae}/2$. While this yields the same threshold for the same growth rate as a special case, we can observe that it is a coincidence. However, this may imply that the condition for EKHI to suppress whistler waves is for the growth rate of the EKHI to be larger than the typical whistler wave frequency---This point needs more verifications.

In this paper, we only consider step function shear flows in inviscid and incompressible plasma. The incompressibility leads to the infinite acoustic wave speed and thus the results obtained in this paper are suitable for low Mach number.  Similar to MHD KHI, the compressibility, non-uniform velocity shear (non-zero vorticity) and density can impact the dispersion relation of EKHI. How compressibility and non-zero vorticity affect the development of EKHI requires further investigations.

\begin{acknowledgements}
H. C. acknowledge partial support by NSF CAREER 2144324 and  a Heliophysics Career award No. 80NSSC19K1106. The authors acknowledge the partial support of an NSF EPSCoR RII-Track-1 Cooperative Agreement OIA-1655280.
\end{acknowledgements}
\newpage 

\begin{thebibliography}{20}%
\makeatletter
\providecommand \@ifxundefined [1]{%
 \@ifx{#1\undefined}
}%
\providecommand \@ifnum [1]{%
 \ifnum #1\expandafter \@firstoftwo
 \else \expandafter \@secondoftwo
 \fi
}%
\providecommand \@ifx [1]{%
 \ifx #1\expandafter \@firstoftwo
 \else \expandafter \@secondoftwo
 \fi
}%
\providecommand \natexlab [1]{#1}%
\providecommand \enquote  [1]{``#1''}%
\providecommand \bibnamefont  [1]{#1}%
\providecommand \bibfnamefont [1]{#1}%
\providecommand \citenamefont [1]{#1}%
\providecommand \href@noop [0]{\@secondoftwo}%
\providecommand \href [0]{\begingroup \@sanitize@url \@href}%
\providecommand \@href[1]{\@@startlink{#1}\@@href}%
\providecommand \@@href[1]{\endgroup#1\@@endlink}%
\providecommand \@sanitize@url [0]{\catcode `\\12\catcode `\$12\catcode
  `\&12\catcode `\#12\catcode `\^12\catcode `\_12\catcode `\%12\relax}%
\providecommand \@@startlink[1]{}%
\providecommand \@@endlink[0]{}%
\providecommand \url  [0]{\begingroup\@sanitize@url \@url }%
\providecommand \@url [1]{\endgroup\@href {#1}{\urlprefix }}%
\providecommand \urlprefix  [0]{URL }%
\providecommand \Eprint [0]{\href }%
\providecommand \doibase [0]{http://dx.doi.org/}%
\providecommand \selectlanguage [0]{\@gobble}%
\providecommand \bibinfo  [0]{\@secondoftwo}%
\providecommand \bibfield  [0]{\@secondoftwo}%
\providecommand \translation [1]{[#1]}%
\providecommand \BibitemOpen [0]{}%
\providecommand \bibitemStop [0]{}%
\providecommand \bibitemNoStop [0]{.\EOS\space}%
\providecommand \EOS [0]{\spacefactor3000\relax}%
\providecommand \BibitemShut  [1]{\csname bibitem#1\endcsname}%
\let\auto@bib@innerbib\@empty
\bibitem [{\citenamefont {{Chandrasekhar}}(1961)}]{chandra61book}%
  \BibitemOpen
  \bibfield  {author} {\bibinfo {author} {\bibfnamefont {S.}~\bibnamefont
  {{Chandrasekhar}}},\ }\href@noop {} {\emph {\bibinfo {title} {{Hydrodynamic
  and hydromagnetic stability}}}}\ (\bibinfo  {publisher} {International Series
  of Monographs on Physics, Oxford: Clarendon},\ \bibinfo {year}
  {1961})\BibitemShut {NoStop}%
\bibitem [{\citenamefont {{D'Angelo}}(1965)}]{dangelo65pof}%
  \BibitemOpen
  \bibfield  {author} {\bibinfo {author} {\bibfnamefont {N.}~\bibnamefont
  {{D'Angelo}}},\ }\bibfield  {title} {\enquote {\bibinfo {title}
  {{Kelvin-Helmholtz Instability in a Fully Ionized Plasma in a Magnetic
  Field}},}\ }\href {\doibase 10.1063/1.1761496} {\bibfield  {journal}
  {\bibinfo  {journal} {Physics of Fluids}\ }\textbf {\bibinfo {volume} {8}},\
  \bibinfo {pages} {1748--1750} (\bibinfo {year} {1965})}\BibitemShut {NoStop}%
\bibitem [{\citenamefont {{Perkins}}\ and\ \citenamefont
  {{Jassby}}(1971)}]{perkins71pof}%
  \BibitemOpen
  \bibfield  {author} {\bibinfo {author} {\bibfnamefont {F.~W.}\ \bibnamefont
  {{Perkins}}}\ and\ \bibinfo {author} {\bibfnamefont {D.~L.}\ \bibnamefont
  {{Jassby}}},\ }\bibfield  {title} {\enquote {\bibinfo {title} {{Velocity
  Shear and Low-Frequency Plasma Instabilities}},}\ }\href {\doibase
  10.1063/1.1693259} {\bibfield  {journal} {\bibinfo  {journal} {Physics of
  Fluids}\ }\textbf {\bibinfo {volume} {14}},\ \bibinfo {pages} {102--115}
  (\bibinfo {year} {1971})}\BibitemShut {NoStop}%
\bibitem [{\citenamefont {{Lau}}\ and\ \citenamefont {{Liu}}(1980)}]{lau80pof}%
  \BibitemOpen
  \bibfield  {author} {\bibinfo {author} {\bibfnamefont {Y.~Y.}\ \bibnamefont
  {{Lau}}}\ and\ \bibinfo {author} {\bibfnamefont {C.~S.}\ \bibnamefont
  {{Liu}}},\ }\bibfield  {title} {\enquote {\bibinfo {title} {{Stability of
  shear flow in a magnetized plasma}},}\ }\href {\doibase 10.1063/1.863100}
  {\bibfield  {journal} {\bibinfo  {journal} {Physics of Fluids}\ }\textbf
  {\bibinfo {volume} {23}},\ \bibinfo {pages} {939--941} (\bibinfo {year}
  {1980})}\BibitemShut {NoStop}%
\bibitem [{\citenamefont {{Miura}}(1982)}]{miura82prl}%
  \BibitemOpen
  \bibfield  {author} {\bibinfo {author} {\bibfnamefont {A.}~\bibnamefont
  {{Miura}}},\ }\bibfield  {title} {\enquote {\bibinfo {title} {{Nonlinear
  evolution of the magnetohydrodynamic Kelvin-Helmholtz instability}},}\ }\href
  {\doibase 10.1103/PhysRevLett.49.779} {\bibfield  {journal} {\bibinfo
  {journal} {\prl}\ }\textbf {\bibinfo {volume} {49}},\ \bibinfo {pages}
  {779--782} (\bibinfo {year} {1982})}\BibitemShut {NoStop}%
\bibitem [{\citenamefont {{Wang}}, \citenamefont {{Pritchett}},\ and\
  \citenamefont {{Ashour-Abdalla}}(1992)}]{wang92pof}%
  \BibitemOpen
  \bibfield  {author} {\bibinfo {author} {\bibfnamefont {Z.}~\bibnamefont
  {{Wang}}}, \bibinfo {author} {\bibfnamefont {P.~L.}\ \bibnamefont
  {{Pritchett}}}, \ and\ \bibinfo {author} {\bibfnamefont {M.}~\bibnamefont
  {{Ashour-Abdalla}}},\ }\bibfield  {title} {\enquote {\bibinfo {title}
  {{Kinetic effects on the velocity-shear-driven instability}},}\ }\href
  {\doibase 10.1063/1.860117} {\bibfield  {journal} {\bibinfo  {journal}
  {Physics of Fluids B}\ }\textbf {\bibinfo {volume} {4}},\ \bibinfo {pages}
  {1092--1101} (\bibinfo {year} {1992})}\BibitemShut {NoStop}%
\bibitem [{\citenamefont {Henri}\ \emph {et~al.}(2013)\citenamefont {Henri},
  \citenamefont {Cerri}, \citenamefont {Califano}, \citenamefont {Pegoraro},
  \citenamefont {Rossi}, \citenamefont {Faganello}, \citenamefont {?ebek},
  \citenamefont {Trávní?ek}, \citenamefont {Hellinger}, \citenamefont
  {Frederiksen}, \citenamefont {Nordlund}, \citenamefont {Markidis},
  \citenamefont {Keppens},\ and\ \citenamefont {Lapenta}}]{henri13pop}%
  \BibitemOpen
  \bibfield  {author} {\bibinfo {author} {\bibfnamefont {P.}~\bibnamefont
  {Henri}}, \bibinfo {author} {\bibfnamefont {S.~S.}\ \bibnamefont {Cerri}},
  \bibinfo {author} {\bibfnamefont {F.}~\bibnamefont {Califano}}, \bibinfo
  {author} {\bibfnamefont {F.}~\bibnamefont {Pegoraro}}, \bibinfo {author}
  {\bibfnamefont {C.}~\bibnamefont {Rossi}}, \bibinfo {author} {\bibfnamefont
  {M.}~\bibnamefont {Faganello}}, \bibinfo {author} {\bibfnamefont
  {O.}~\bibnamefont {?ebek}}, \bibinfo {author} {\bibfnamefont {P.~M.}\
  \bibnamefont {Trávní?ek}}, \bibinfo {author} {\bibfnamefont
  {P.}~\bibnamefont {Hellinger}}, \bibinfo {author} {\bibfnamefont {J.~T.}\
  \bibnamefont {Frederiksen}}, \bibinfo {author} {\bibfnamefont
  {A.}~\bibnamefont {Nordlund}}, \bibinfo {author} {\bibfnamefont
  {S.}~\bibnamefont {Markidis}}, \bibinfo {author} {\bibfnamefont
  {R.}~\bibnamefont {Keppens}}, \ and\ \bibinfo {author} {\bibfnamefont
  {G.}~\bibnamefont {Lapenta}},\ }\bibfield  {title} {\enquote {\bibinfo
  {title} {Nonlinear evolution of the magnetized kelvin-helmholtz instability:
  From fluid to kinetic modeling},}\ }\href {\doibase 10.1063/1.4826214}
  {\bibfield  {journal} {\bibinfo  {journal} {Physics of Plasmas}\ }\textbf
  {\bibinfo {volume} {20}},\ \bibinfo {pages} {102118} (\bibinfo {year}
  {2013})}\BibitemShut {NoStop}%
\bibitem [{\citenamefont {Liljeblad}\ \emph {et~al.}(2014)\citenamefont
  {Liljeblad}, \citenamefont {Sundberg}, \citenamefont {Karlsson},\ and\
  \citenamefont {Kullen}}]{lilje14jgr}%
  \BibitemOpen
  \bibfield  {author} {\bibinfo {author} {\bibfnamefont {E.}~\bibnamefont
  {Liljeblad}}, \bibinfo {author} {\bibfnamefont {T.}~\bibnamefont {Sundberg}},
  \bibinfo {author} {\bibfnamefont {T.}~\bibnamefont {Karlsson}}, \ and\
  \bibinfo {author} {\bibfnamefont {A.}~\bibnamefont {Kullen}},\ }\bibfield
  {title} {\enquote {\bibinfo {title} {Statistical investigation of
  kelvin-helmholtz waves at the magnetopause of mercury},}\ }\href {\doibase
  https://doi.org/10.1002/2014JA020614} {\bibfield  {journal} {\bibinfo
  {journal} {Journal of Geophysical Research: Space Physics}\ }\textbf
  {\bibinfo {volume} {119}},\ \bibinfo {pages} {9670--9683} (\bibinfo {year}
  {2014})}\BibitemShut {NoStop}%
\bibitem [{\citenamefont {{Huang}}\ \emph {et~al.}(2017)\citenamefont
  {{Huang}}, \citenamefont {{Lu}}, \citenamefont {{Wang}}, \citenamefont
  {{Guo}}, \citenamefont {{Wu}}, \citenamefont {{Lu}},\ and\ \citenamefont
  {{Wang}}}]{huangC17apj}%
  \BibitemOpen
  \bibfield  {author} {\bibinfo {author} {\bibfnamefont {C.}~\bibnamefont
  {{Huang}}}, \bibinfo {author} {\bibfnamefont {Q.}~\bibnamefont {{Lu}}},
  \bibinfo {author} {\bibfnamefont {R.}~\bibnamefont {{Wang}}}, \bibinfo
  {author} {\bibfnamefont {F.}~\bibnamefont {{Guo}}}, \bibinfo {author}
  {\bibfnamefont {M.}~\bibnamefont {{Wu}}}, \bibinfo {author} {\bibfnamefont
  {S.}~\bibnamefont {{Lu}}}, \ and\ \bibinfo {author} {\bibfnamefont
  {S.}~\bibnamefont {{Wang}}},\ }\bibfield  {title} {\enquote {\bibinfo {title}
  {{Development of Turbulent Magnetic Reconnection in a Magnetic Island}},}\
  }\href {\doibase 10.3847/1538-4357/835/2/245} {\bibfield  {journal} {\bibinfo
   {journal} {\apj}\ }\textbf {\bibinfo {volume} {835}},\ \bibinfo {eid} {245}
  (\bibinfo {year} {2017})}\BibitemShut {NoStop}%
\bibitem [{\citenamefont {Zhong}\ \emph {et~al.}(2018)\citenamefont {Zhong},
  \citenamefont {Tang}, \citenamefont {Zhou}, \citenamefont {Deng},
  \citenamefont {Pang}, \citenamefont {Paterson}, \citenamefont {Giles},
  \citenamefont {Burch}, \citenamefont {Tobert}, \citenamefont {Ergun},
  \citenamefont {Khotyaintsev},\ and\ \citenamefont {Lindquist}}]{zhong18prl}%
  \BibitemOpen
  \bibfield  {author} {\bibinfo {author} {\bibfnamefont {Z.~H.}\ \bibnamefont
  {Zhong}}, \bibinfo {author} {\bibfnamefont {R.~X.}\ \bibnamefont {Tang}},
  \bibinfo {author} {\bibfnamefont {M.}~\bibnamefont {Zhou}}, \bibinfo {author}
  {\bibfnamefont {X.~H.}\ \bibnamefont {Deng}}, \bibinfo {author}
  {\bibfnamefont {Y.}~\bibnamefont {Pang}}, \bibinfo {author} {\bibfnamefont
  {W.~R.}\ \bibnamefont {Paterson}}, \bibinfo {author} {\bibfnamefont {B.~L.}\
  \bibnamefont {Giles}}, \bibinfo {author} {\bibfnamefont {J.~L.}\ \bibnamefont
  {Burch}}, \bibinfo {author} {\bibfnamefont {R.~B.}\ \bibnamefont {Tobert}},
  \bibinfo {author} {\bibfnamefont {R.~E.}\ \bibnamefont {Ergun}}, \bibinfo
  {author} {\bibfnamefont {Y.~V.}\ \bibnamefont {Khotyaintsev}}, \ and\
  \bibinfo {author} {\bibfnamefont {P.-A.}\ \bibnamefont {Lindquist}},\
  }\bibfield  {title} {\enquote {\bibinfo {title} {Evidence for secondary flux
  rope generated by the electron kelvin-helmholtz instability in a magnetic
  reconnection diffusion region},}\ }\href {\doibase
  10.1103/PhysRevLett.120.075101} {\bibfield  {journal} {\bibinfo  {journal}
  {Phys. Rev. Lett.}\ }\textbf {\bibinfo {volume} {120}},\ \bibinfo {pages}
  {075101} (\bibinfo {year} {2018})}\BibitemShut {NoStop}%
\bibitem [{\citenamefont {Hwang}\ \emph {et~al.}(2019)\citenamefont {Hwang},
  \citenamefont {Choi}, \citenamefont {Dokgo}, \citenamefont {Burch},
  \citenamefont {Sibeck}, \citenamefont {Giles}, \citenamefont {Goldstein},
  \citenamefont {Paterson}, \citenamefont {Pollock}, \citenamefont {Shi},
  \citenamefont {Fu}, \citenamefont {Hasegawa}, \citenamefont {Gershman},
  \citenamefont {Khotyaintsev}, \citenamefont {Torbert}, \citenamefont {Ergun},
  \citenamefont {Dorelli}, \citenamefont {Avanov}, \citenamefont {Russell},\
  and\ \citenamefont {Strangeway}}]{hwang19grl}%
  \BibitemOpen
  \bibfield  {author} {\bibinfo {author} {\bibfnamefont {K.-J.}\ \bibnamefont
  {Hwang}}, \bibinfo {author} {\bibfnamefont {E.}~\bibnamefont {Choi}},
  \bibinfo {author} {\bibfnamefont {K.}~\bibnamefont {Dokgo}}, \bibinfo
  {author} {\bibfnamefont {J.~L.}\ \bibnamefont {Burch}}, \bibinfo {author}
  {\bibfnamefont {D.~G.}\ \bibnamefont {Sibeck}}, \bibinfo {author}
  {\bibfnamefont {B.~L.}\ \bibnamefont {Giles}}, \bibinfo {author}
  {\bibfnamefont {M.~L.}\ \bibnamefont {Goldstein}}, \bibinfo {author}
  {\bibfnamefont {W.~R.}\ \bibnamefont {Paterson}}, \bibinfo {author}
  {\bibfnamefont {C.~J.}\ \bibnamefont {Pollock}}, \bibinfo {author}
  {\bibfnamefont {Q.~Q.}\ \bibnamefont {Shi}}, \bibinfo {author} {\bibfnamefont
  {H.}~\bibnamefont {Fu}}, \bibinfo {author} {\bibfnamefont {H.}~\bibnamefont
  {Hasegawa}}, \bibinfo {author} {\bibfnamefont {D.~J.}\ \bibnamefont
  {Gershman}}, \bibinfo {author} {\bibfnamefont {Y.}~\bibnamefont
  {Khotyaintsev}}, \bibinfo {author} {\bibfnamefont {R.~B.}\ \bibnamefont
  {Torbert}}, \bibinfo {author} {\bibfnamefont {R.~E.}\ \bibnamefont {Ergun}},
  \bibinfo {author} {\bibfnamefont {J.~C.}\ \bibnamefont {Dorelli}}, \bibinfo
  {author} {\bibfnamefont {L.}~\bibnamefont {Avanov}}, \bibinfo {author}
  {\bibfnamefont {C.~T.}\ \bibnamefont {Russell}}, \ and\ \bibinfo {author}
  {\bibfnamefont {R.~J.}\ \bibnamefont {Strangeway}},\ }\bibfield  {title}
  {\enquote {\bibinfo {title} {Electron vorticity indicative of the electron
  diffusion region of magnetic reconnection},}\ }\href {\doibase
  https://doi.org/10.1029/2019GL082710} {\bibfield  {journal} {\bibinfo
  {journal} {Geophysical Research Letters}\ }\textbf {\bibinfo {volume} {46}},\
  \bibinfo {pages} {6287--6296} (\bibinfo {year} {2019})},\ \Eprint
  {http://arxiv.org/abs/https://agupubs.onlinelibrary.wiley.com/doi/pdf/10.1029/2019GL082710}
  {https://agupubs.onlinelibrary.wiley.com/doi/pdf/10.1029/2019GL082710}
  \BibitemShut {NoStop}%
\bibitem [{\citenamefont {{Che}}, \citenamefont {{Zank}},\ and\ \citenamefont
  {{Benz}}(2021)}]{che21apjb}%
  \BibitemOpen
  \bibfield  {author} {\bibinfo {author} {\bibfnamefont {H.}~\bibnamefont
  {{Che}}}, \bibinfo {author} {\bibfnamefont {G.~P.}\ \bibnamefont {{Zank}}}, \
  and\ \bibinfo {author} {\bibfnamefont {A.~O.}\ \bibnamefont {{Benz}}},\
  }\bibfield  {title} {\enquote {\bibinfo {title} {{Ion Acceleration and the
  Development of a Power-law Energy Spectrum in Magnetic Reconnection}},}\
  }\href {\doibase 10.3847/1538-4357/ac1fe7} {\bibfield  {journal} {\bibinfo
  {journal} {\apj}\ }\textbf {\bibinfo {volume} {921}},\ \bibinfo {eid} {135}
  (\bibinfo {year} {2021})}\BibitemShut {NoStop}%
\bibitem [{\citenamefont {Che}\ \emph {et~al.}(2021)\citenamefont {Che},
  \citenamefont {Zank}, \citenamefont {Benz}, \citenamefont {Tang},\ and\
  \citenamefont {Crawford}}]{che21apja}%
  \BibitemOpen
  \bibfield  {author} {\bibinfo {author} {\bibfnamefont {H.}~\bibnamefont
  {Che}}, \bibinfo {author} {\bibfnamefont {G.~P.}\ \bibnamefont {Zank}},
  \bibinfo {author} {\bibfnamefont {A.~O.}\ \bibnamefont {Benz}}, \bibinfo
  {author} {\bibfnamefont {B.}~\bibnamefont {Tang}}, \ and\ \bibinfo {author}
  {\bibfnamefont {C.}~\bibnamefont {Crawford}},\ }\bibfield  {title} {\enquote
  {\bibinfo {title} {The formation of electron outflow jets with power-law
  energy distribution in guide-field magnetic reconnection},}\ }\href {\doibase
  10.3847/1538-4357/abcf29} {\bibfield  {journal} {\bibinfo  {journal} {\apj}\
  }\textbf {\bibinfo {volume} {908}},\ \bibinfo {pages} {72} (\bibinfo {year}
  {2021})}\BibitemShut {NoStop}%
\bibitem [{\citenamefont {{Delamere}}\ \emph {et~al.}(2021)\citenamefont
  {{Delamere}}, \citenamefont {{Ng}}, \citenamefont {{Damiano}}, \citenamefont
  {{Neupane}}, \citenamefont {{Johnson}}, \citenamefont {{Burkholder}},
  \citenamefont {{Ma}},\ and\ \citenamefont {{Nykyri}}}]{dela21jgr}%
  \BibitemOpen
  \bibfield  {author} {\bibinfo {author} {\bibfnamefont {P.~A.}\ \bibnamefont
  {{Delamere}}}, \bibinfo {author} {\bibfnamefont {C.~S.}\ \bibnamefont
  {{Ng}}}, \bibinfo {author} {\bibfnamefont {P.~A.}\ \bibnamefont {{Damiano}}},
  \bibinfo {author} {\bibfnamefont {B.~R.}\ \bibnamefont {{Neupane}}}, \bibinfo
  {author} {\bibfnamefont {J.~R.}\ \bibnamefont {{Johnson}}}, \bibinfo {author}
  {\bibfnamefont {B.}~\bibnamefont {{Burkholder}}}, \bibinfo {author}
  {\bibfnamefont {X.}~\bibnamefont {{Ma}}}, \ and\ \bibinfo {author}
  {\bibfnamefont {K.}~\bibnamefont {{Nykyri}}},\ }\bibfield  {title} {\enquote
  {\bibinfo {title} {{Kelvin-Helmholtz Related Turbulent Heating at Saturn's
  Magnetopause Boundary}},}\ }\href {\doibase 10.1029/2020JA028479} {\bibfield
  {journal} {\bibinfo  {journal} {Journal of Geophysical Research (Space
  Physics)}\ }\textbf {\bibinfo {volume} {126}},\ \bibinfo {eid} {e28479}
  (\bibinfo {year} {2021})}\BibitemShut {NoStop}%
\bibitem [{\citenamefont {{Che}}\ and\ \citenamefont
  {{Zank}}(2020)}]{che20apj}%
  \BibitemOpen
  \bibfield  {author} {\bibinfo {author} {\bibfnamefont {H.}~\bibnamefont
  {{Che}}}\ and\ \bibinfo {author} {\bibfnamefont {G.~P.}\ \bibnamefont
  {{Zank}}},\ }\bibfield  {title} {\enquote {\bibinfo {title} {{Electron
  Acceleration from Expanding Magnetic Vortices During Reconnection with a
  Guide Field}},}\ }\href {\doibase 10.3847/1538-4357/ab5d3b} {\bibfield
  {journal} {\bibinfo  {journal} {\apj}\ }\textbf {\bibinfo {volume} {889}},\
  \bibinfo {eid} {11} (\bibinfo {year} {2020})}\BibitemShut {NoStop}%
\bibitem [{\citenamefont {{Li}}\ \emph {et~al.}(2022)\citenamefont {{Li}},
  \citenamefont {{Wang}}, \citenamefont {{Lu}}, \citenamefont {{Russell}},
  \citenamefont {{Lu}}, \citenamefont {{Cohen}}, \citenamefont {{Ergun}},\ and\
  \citenamefont {{Wang}}}]{li22natcommu}%
  \BibitemOpen
  \bibfield  {author} {\bibinfo {author} {\bibfnamefont {X.}~\bibnamefont
  {{Li}}}, \bibinfo {author} {\bibfnamefont {R.}~\bibnamefont {{Wang}}},
  \bibinfo {author} {\bibfnamefont {Q.}~\bibnamefont {{Lu}}}, \bibinfo {author}
  {\bibfnamefont {C.~T.}\ \bibnamefont {{Russell}}}, \bibinfo {author}
  {\bibfnamefont {S.}~\bibnamefont {{Lu}}}, \bibinfo {author} {\bibfnamefont
  {I.~J.}\ \bibnamefont {{Cohen}}}, \bibinfo {author} {\bibfnamefont {R.~E.}\
  \bibnamefont {{Ergun}}}, \ and\ \bibinfo {author} {\bibfnamefont
  {S.}~\bibnamefont {{Wang}}},\ }\bibfield  {title} {\enquote {\bibinfo {title}
  {{Three-dimensional network of filamentary currents and super-thermal
  electrons during magnetotail magnetic reconnection}},}\ }\href {\doibase
  10.1038/s41467-022-31025-9} {\bibfield  {journal} {\bibinfo  {journal}
  {Nature Communications}\ }\textbf {\bibinfo {volume} {13}},\ \bibinfo {eid}
  {3241} (\bibinfo {year} {2022})}\BibitemShut {NoStop}%
\bibitem [{\citenamefont {{Wilder}}\ \emph {et~al.}(2016)\citenamefont
  {{Wilder}}, \citenamefont {{Ergun}}, \citenamefont {{Goodrich}},
  \citenamefont {{Goldman}}, \citenamefont {{Newman}}, \citenamefont
  {{Malaspina}}, \citenamefont {{Jaynes}}, \citenamefont {{Schwartz}},
  \citenamefont {{Trattner}}, \citenamefont {{Burch}}, \citenamefont
  {{Argall}}, \citenamefont {{Torbert}}, \citenamefont {{Lindqvist}},
  \citenamefont {{Marklund}}, \citenamefont {{Le Contel}}, \citenamefont
  {{Mirioni}}, \citenamefont {{Khotyaintsev}}, \citenamefont {{Strangeway}},
  \citenamefont {{Russell}}, \citenamefont {{Pollock}}, \citenamefont
  {{Giles}}, \citenamefont {{Plaschke}}, \citenamefont {{Magnes}},
  \citenamefont {{Eriksson}}, \citenamefont {{Stawarz}}, \citenamefont
  {{Sturner}},\ and\ \citenamefont {{Holmes}}}]{wilder16grl}%
  \BibitemOpen
  \bibfield  {author} {\bibinfo {author} {\bibfnamefont {F.~D.}\ \bibnamefont
  {{Wilder}}}, \bibinfo {author} {\bibfnamefont {R.~E.}\ \bibnamefont
  {{Ergun}}}, \bibinfo {author} {\bibfnamefont {K.~A.}\ \bibnamefont
  {{Goodrich}}}, \bibinfo {author} {\bibfnamefont {M.~V.}\ \bibnamefont
  {{Goldman}}}, \bibinfo {author} {\bibfnamefont {D.~L.}\ \bibnamefont
  {{Newman}}}, \bibinfo {author} {\bibfnamefont {D.~M.}\ \bibnamefont
  {{Malaspina}}}, \bibinfo {author} {\bibfnamefont {A.~N.}\ \bibnamefont
  {{Jaynes}}}, \bibinfo {author} {\bibfnamefont {S.~J.}\ \bibnamefont
  {{Schwartz}}}, \bibinfo {author} {\bibfnamefont {K.~J.}\ \bibnamefont
  {{Trattner}}}, \bibinfo {author} {\bibfnamefont {J.~L.}\ \bibnamefont
  {{Burch}}}, \bibinfo {author} {\bibfnamefont {M.~R.}\ \bibnamefont
  {{Argall}}}, \bibinfo {author} {\bibfnamefont {R.~B.}\ \bibnamefont
  {{Torbert}}}, \bibinfo {author} {\bibfnamefont {P.-A.}\ \bibnamefont
  {{Lindqvist}}}, \bibinfo {author} {\bibfnamefont {G.}~\bibnamefont
  {{Marklund}}}, \bibinfo {author} {\bibfnamefont {O.}~\bibnamefont {{Le
  Contel}}}, \bibinfo {author} {\bibfnamefont {L.}~\bibnamefont {{Mirioni}}},
  \bibinfo {author} {\bibfnamefont {Y.~V.}\ \bibnamefont {{Khotyaintsev}}},
  \bibinfo {author} {\bibfnamefont {R.~J.}\ \bibnamefont {{Strangeway}}},
  \bibinfo {author} {\bibfnamefont {C.~T.}\ \bibnamefont {{Russell}}}, \bibinfo
  {author} {\bibfnamefont {C.~J.}\ \bibnamefont {{Pollock}}}, \bibinfo {author}
  {\bibfnamefont {B.~L.}\ \bibnamefont {{Giles}}}, \bibinfo {author}
  {\bibfnamefont {F.}~\bibnamefont {{Plaschke}}}, \bibinfo {author}
  {\bibfnamefont {W.}~\bibnamefont {{Magnes}}}, \bibinfo {author}
  {\bibfnamefont {S.}~\bibnamefont {{Eriksson}}}, \bibinfo {author}
  {\bibfnamefont {J.~E.}\ \bibnamefont {{Stawarz}}}, \bibinfo {author}
  {\bibfnamefont {A.~P.}\ \bibnamefont {{Sturner}}}, \ and\ \bibinfo {author}
  {\bibfnamefont {J.~C.}\ \bibnamefont {{Holmes}}},\ }\bibfield  {title}
  {\enquote {\bibinfo {title} {{Observations of whistler mode waves with
  nonlinear parallel electric fields near the dayside magnetic reconnection
  separatrix by the Magnetospheric Multiscale mission}},}\ }\href {\doibase
  10.1002/2016GL069473} {\bibfield  {journal} {\bibinfo  {journal} {\grl}\
  }\textbf {\bibinfo {volume} {43}},\ \bibinfo {pages} {5909--5917} (\bibinfo
  {year} {2016})}\BibitemShut {NoStop}%
\bibitem [{\citenamefont {Wilder}\ \emph {et~al.}(2020)\citenamefont {Wilder},
  \citenamefont {Schwartz}, \citenamefont {Ergun}, \citenamefont {Eriksson},
  \citenamefont {Ahmadi}, \citenamefont {Chasapis}, \citenamefont {Newman},
  \citenamefont {Burch}, \citenamefont {Torbert}, \citenamefont {Strangeway},\
  and\ \citenamefont {Giles}}]{wilder20grl}%
  \BibitemOpen
  \bibfield  {author} {\bibinfo {author} {\bibfnamefont {F.~D.}\ \bibnamefont
  {Wilder}}, \bibinfo {author} {\bibfnamefont {S.~J.}\ \bibnamefont
  {Schwartz}}, \bibinfo {author} {\bibfnamefont {R.~E.}\ \bibnamefont {Ergun}},
  \bibinfo {author} {\bibfnamefont {S.}~\bibnamefont {Eriksson}}, \bibinfo
  {author} {\bibfnamefont {N.}~\bibnamefont {Ahmadi}}, \bibinfo {author}
  {\bibfnamefont {A.}~\bibnamefont {Chasapis}}, \bibinfo {author}
  {\bibfnamefont {D.~L.}\ \bibnamefont {Newman}}, \bibinfo {author}
  {\bibfnamefont {J.~L.}\ \bibnamefont {Burch}}, \bibinfo {author}
  {\bibfnamefont {R.~B.}\ \bibnamefont {Torbert}}, \bibinfo {author}
  {\bibfnamefont {R.~J.}\ \bibnamefont {Strangeway}}, \ and\ \bibinfo {author}
  {\bibfnamefont {B.~L.}\ \bibnamefont {Giles}},\ }\bibfield  {title} {\enquote
  {\bibinfo {title} {Parallel electrostatic waves associated with turbulent
  plasma mixing in the kelvin-helmholtz instability},}\ }\href {\doibase
  https://doi.org/10.1029/2020GL087837} {\bibfield  {journal} {\bibinfo
  {journal} {Geophysical Research Letters}\ }\textbf {\bibinfo {volume} {47}},\
  \bibinfo {pages} {e2020GL087837} (\bibinfo {year} {2020})},\ \bibinfo {note}
  {e2020GL087837 2020GL087837}\BibitemShut {NoStop}%
\bibitem [{\citenamefont {{Birn}}\ \emph {et~al.}(2001)\citenamefont {{Birn}},
  \citenamefont {{Drake}}, \citenamefont {{Shay}}, \citenamefont {{Rogers}},
  \citenamefont {{Denton}}, \citenamefont {{Hesse}}, \citenamefont
  {{Kuznetsova}}, \citenamefont {{Ma}}, \citenamefont {{Bhattacharjee}},
  \citenamefont {{Otto}},\ and\ \citenamefont {{Pritchett}}}]{birn01jgrb}%
  \BibitemOpen
  \bibfield  {author} {\bibinfo {author} {\bibfnamefont {J.}~\bibnamefont
  {{Birn}}}, \bibinfo {author} {\bibfnamefont {J.~F.}\ \bibnamefont {{Drake}}},
  \bibinfo {author} {\bibfnamefont {M.~A.}\ \bibnamefont {{Shay}}}, \bibinfo
  {author} {\bibfnamefont {B.~N.}\ \bibnamefont {{Rogers}}}, \bibinfo {author}
  {\bibfnamefont {R.~E.}\ \bibnamefont {{Denton}}}, \bibinfo {author}
  {\bibfnamefont {M.}~\bibnamefont {{Hesse}}}, \bibinfo {author} {\bibfnamefont
  {M.}~\bibnamefont {{Kuznetsova}}}, \bibinfo {author} {\bibfnamefont {Z.~W.}\
  \bibnamefont {{Ma}}}, \bibinfo {author} {\bibfnamefont {A.}~\bibnamefont
  {{Bhattacharjee}}}, \bibinfo {author} {\bibfnamefont {A.}~\bibnamefont
  {{Otto}}}, \ and\ \bibinfo {author} {\bibfnamefont {P.~L.}\ \bibnamefont
  {{Pritchett}}},\ }\bibfield  {title} {\enquote {\bibinfo {title} {{Geospace
  Environmental Modeling (GEM) magnetic reconnection challenge}},}\ }\href
  {\doibase 10.1029/1999JA900449} {\bibfield  {journal} {\bibinfo  {journal}
  {\jgr}\ }\textbf {\bibinfo {volume} {106}},\ \bibinfo {pages} {3715--3720}
  (\bibinfo {year} {2001})}\BibitemShut {NoStop}%
\bibitem [{\citenamefont {{Fermo}}, \citenamefont {{Drake}},\ and\
  \citenamefont {{Swisdak}}(2012)}]{Fermo12prl}%
  \BibitemOpen
  \bibfield  {author} {\bibinfo {author} {\bibfnamefont {R.~L.}\ \bibnamefont
  {{Fermo}}}, \bibinfo {author} {\bibfnamefont {J.~F.}\ \bibnamefont
  {{Drake}}}, \ and\ \bibinfo {author} {\bibfnamefont {M.}~\bibnamefont
  {{Swisdak}}},\ }\bibfield  {title} {\enquote {\bibinfo {title} {{Secondary
  Magnetic Islands Generated by the Kelvin-Helmholtz Instability in a
  Reconnecting Current Sheet}},}\ }\href {\doibase
  10.1103/PhysRevLett.108.255005} {\bibfield  {journal} {\bibinfo  {journal}
  {Physical Review Letters}\ }\textbf {\bibinfo {volume} {108}},\ \bibinfo
  {eid} {255005} (\bibinfo {year} {2012})}\BibitemShut {NoStop}%
\end{thebibliography}

%

\end{document}